\documentstyle[11pt, paspconf, epsf]{article}
\begin{document}

\title{Halo profiles from weakly nonlinear initial conditions}
\author{Ewa L. \L okas}
\affil{Copernicus Astronomical Center, Bartycka 18, 00-716 Warsaw, 
Poland}
\begin{abstract}
Using weakly nonlinear conditional PDF for the density field around an 
overdense region we find that the expected density contrast around a peak 
is always smaller while its rms fluctuation larger than in the linear 
case. We apply these results to the spherical model of collapse as 
developed by Hoffman \& Shaham (1985) and find that in the case of open 
Universe the effect of weakly nonlinear interactions is to decrease the 
scale from which a peak gathers mass and the mass itself as well as to 
steepen the final profile of the virialized protoobject. 
\end{abstract}

\section{Introduction}

The purpose of this work was to find a generalization of the spherical 
infall model as developed by Hoffman \& Shaham (1985, hereafter HS) to the 
case of density peaks collapsing in the weakly nonlinear background. 
One of the key assumptions underlying the calculations of HS was that the 
matter influenced by the peak collapses onto it undisturbed by the 
background. This is equivalent to the statement that the peak identified 
with some resolution (smoothing) scale collapses while the surrounding 
density field is still linear i.e. its rms fluctuation at this scale is 
much smaller than unity. The example of the Virgo supercluster (which has 
not yet collapsed) shows that this is not always the case: the rms 
fluctuation at the scale of the supercluster is well in the weakly 
nonlinear regime.

In this generalization we hope to account properly for the weakly 
nonlinear transition between the linear and strongly nonlinear phase of 
the evolution of the perturbation which lacked in the approach of HS. It 
involves constructing the weakly nonlinear probability distribution 
function (PDF) of density around an overdense region. The mean density 
obtained from this weakly nonlinear PDF is then taken as the initial 
condition for spherical collapse. 

The reliability of this approach rests on the assumption that the 
influence of the neighbouring fluctuations can be restricted to the weakly 
nonlinear phase with its only outcome in the form of a changed `initial' 
density profile which then evolves independently of surroundings, 
according to the spherical model.
 
\section{Conditional probability distribution around a peak}

\begin{figure} 
\begin{center}
    \leavevmode
    \epsfxsize=8cm
    \epsfbox[0 200 550 710]{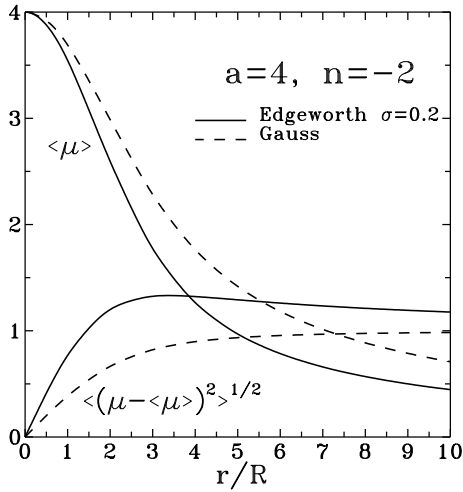}
\end{center}
    \caption{The expected normalized density contrast $\langle\mu\rangle$ 
    and the dispersion $\langle(\mu - 
    \langle\mu\rangle)^{2}\rangle^{1/2}$ for the Edgeworth and Gaussian 
    conditional distributions as a function of the distance from the peak.} 
\label{fig1}
\end{figure}

We consider density contrast field which initially has Gaussian
distribution with zero mean. The field measured at a given point is
denoted by $\delta$ while the one measured at the distance $r$
from the first point is called $\gamma$. We smooth the fields with 
Gaussian filters of the same scale $R$ so their variances are equal 
$\langle\delta^{2}\rangle=\langle\gamma^{2}\rangle=\sigma^{2}$.

We generalize the joint probability distribution function for density 
contrast field in the Gaussian case to the case when the density fields 
are weakly nonlinear and therefore departing from Gaussianity. If the rms 
fluctuations of the fields are small (below unity) the fields can be 
expanded around their linear values $\delta_{1}$ and $\gamma_{1}$. The 
resulting two-point PDF takes the form of the bivariate Edgeworth series 
(\L okas 1997). 

Let us now suppose that the value of one variable is 
known: $\gamma = a \sigma$, i.e. at $r=0$ we have a region of $a$ 
standard deviations. The conditional probability distribution of the 
normalized density $\mu = \delta/\sigma$ at distance $r$ from this region 
is well characterized by its lowest order moments with respect to the mean 
value, $\langle\mu\rangle$. In the Gaussian case we have 
$\langle\mu\rangle_{G} = \varrho a$ and $\langle(\mu - 
\langle\mu\rangle)^{2}\rangle_{G} = 1 - \varrho^{2}$, while in the weakly 
nonlinear approximation we get (\L okas 1997) 
\begin{eqnarray} 
    \langle\mu\rangle &=& \varrho \ a + \frac{\sigma}{2} (a^{2} - 1) 
    (S_{12} - \varrho S_{3}) \label{l27}\\ \langle(\mu - 
    \langle\mu\rangle)^{2}\rangle &=& 1 - \varrho^{2} + \sigma \ a \ 
    (S_{12} - 2 \varrho S_{12} + \varrho^{2} S_{3}) \label{l28} 
\end{eqnarray} 
where $\varrho = \xi_{R}(r) / \sigma^{2}$ is the 
correlation coefficient and $S_3$ and $S_{12}$ are the one-point and 
two-point skewness parameters respectively. 

It turns out that for Gaussian smoothing, scale-free power spectra 
$P(k)=Ck^n$ with indices $n=-2,-1.5,-1$ we have $S_{12} - \varrho S_{3} < 
0$. This proves that according to equation (\ref{l27}) for 
$a>1$ (peaks) the correction to the mean normalized density contrast with 
respect to the Gaussian value is always negative. Figure~\ref{fig1} shows 
that the effect of weakly nonlinear interactions is to decrease the 
expected density around an overdense region. In the case of the variance 
(Figure~\ref{fig1}) the weakly nonlinear corrections work in the opposite 
direction: their effect is to increase the value of the variance (or 
dispersion) with respect to the linear case, because for the power spectra 
considered here we always have $S_{12} - 2 \varrho S_{12} + \varrho^{2} 
S_{3} > 0$.

It is clear that a local density peak that rises significantly 
above the noise should gravitationally dominate its surroundings out to 
some distance. A reasonable measure of the distance, up to which a 
coherent structure around the peak is expected, is the scale $r_{coh}$ at 
which $\langle\mu\rangle = \langle(\mu - 
\langle\mu\rangle)^{2}\rangle^{1/2}$. Figure~\ref{fig1} shows that 
the coherence scale in weakly nonlinear regime is decreased with respect 
to the linear case.

\section{Application to spherical collapse}

The dynamical evolution of matter at the distance $c_{i}=r_{i}/R$ from the
peak is determined by the mean cumulative density perturbation within
$c_{i}$ which in weakly nonlinear approximation preserves its slope 
$c_{i}^{-(n+3)}$ for large $c_{i}$ but the height of the peak instead of 
the linear value $a$ can be approximated by 
$a_{eff} = a [1 - \sigma s (a^{2}-1)/(2 a)]$,
where $s$ is of the order of unity.
If we now assume that the matter within
distance $c_{i}$ from the peak collapses undisturbed onto the peak, the
spherical model can be applied. Following the calculations of HS with the 
initial conditions settled by the Edgeworth approximation instead of the 
Gaussian one we find the final density profile to be given by 
\begin{equation}  \label{l43}
    \rho(c) \propto 
    \frac{[(c_{0}/c_{i})^{n+3} - 1]^{4}}{(n+4)(c_{0}/c_{i})^{n+3} - 1},
    \ \ \ \ \ \ \ c_{i}/c \propto (c_{0}/c_{i})^{n+3} - 1
\end{equation}
which has the same form as that of HS (so that the limiting 
cases $\rho(c) \propto (c_{0}/c )^{3(n+3)/(n+4)}$ for $c_{i} \ll c_{0}$
and $\rho(c) \propto (c_{0}/c)^{4}$ for $c_{i} \le c_{0}$ are preserved)
but the scale from which the peak gathers mass (the distance to the shell 
of zero energy) is decreased with respect to the linear case
\begin{equation}  \label{l39}
    c_{0} = c_{0, G} \left[1-\frac{\sigma s (a^2 - 1)}{2 a (n+3)} \right].
\end{equation}
The value of $c_{0}$ determines the slope of the profile (\ref{l43}), the 
shorter $c_{0}$ the steeper is the density profile. The value of $c_{0}$ 
is defined so that in flat Universe it is infinite therefore in this case 
the weakly nonlinear corrections do not affect the final profile. In the 
open Universe, however, the correction to $c_{0}$ is significant and leads 
to the steepening of the final profile and decreasing the mass bound to 
the peak.

\end{document}